\documentclass[doublecol]{epl2} 
\usepackage{graphicx}

\title{{\bf Quantum computing for fluids: where do we stand?}}

\author{Sauro Succi \inst{1} \and W. Itani \inst{2} \and K. Sreenivasan \inst{2} \and
R. Steijl \inst{3}} 
\institute{
\inst{1} Fondazione Istituto Italiano di Tecnologia,\\
Center for Life Nano-Neuroscience at la Sapienza, Viale Regina Elena 291, 00161 Roma, Italy\\
\inst{2} Tandon School of Engineering, New York University,\\
Brooklyn, New York, NY 11201, USA\\
\inst{3} James Watt School of Engineering, University of Glasgow, Glasgow G12 8QQ, UK
}

\abstract{
We present a pedagogical introduction to the current state of 
quantum  computing algorithms for the simulation of classical fluids.
Different strategies, along with their potential merits and liabilities, 
are discussed and commented on.
}

\begin{document}

\maketitle

\section{Introduction}

Quantum computing is one of the most uproaring topics of modern  
science, holding promises of spectacular applications far beyond 
the reach of classical electronic computers, at least for selected 
applications\cite{QC}.  

The manifesto of quantum computing can be traced back to Richard Feynman's 
epoch-making paper, in which he famously observed that physics "ain't classical", hence 
it ought to be simulated on quantum computers \cite{FEY}.

\begin{figure}
\centering
\includegraphics[scale=0.3]{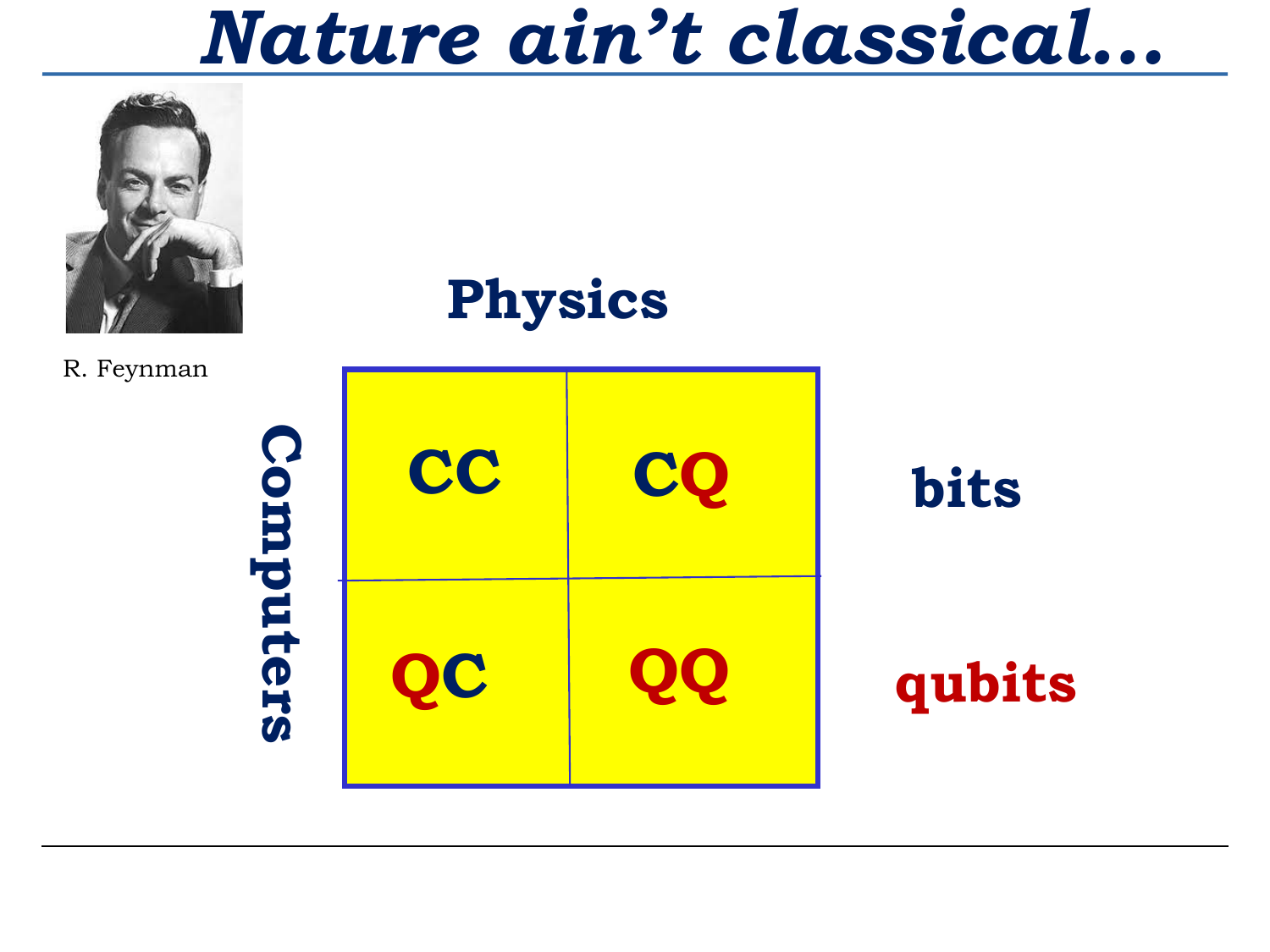}
\caption{The four quadrants in the Physics-Computing plane. 
The CC and CQ quadrants are the mainstay of current simulation work.
QQ is the quadrant invoked by Feynman, and QC is the quadrant addressed
by the present Perspective.
}
\label{fig_CCQQ}
\end{figure}

Following Feynman's observations, early theoretical work on quantum computing 
was performed in the 1980s, e.g. Deutsch's work on the link between quantum 
theory, universal quantum computers and the Church–Turing principle\cite{Deutsch1985}. 
Then, with the publication of Shor's algorithm for integer factoring and Grover's 
search algorithm in the middle of the 1990s, the research area gathered significant 
momentum in terms of theoretical work and quantum computing hardware as well.
The research area of quantum computing has continued to grow ever since  
\cite{Preskill2018,IBM,Bravyi2022}. 
In terms of applications for quantum computers, the simulation of quantum 
many-body systems has received the most attention, due to its scientific and industrial 
applications, as well as the relatively close link with quantum hardware, as per 
Feynman's original proposal.
In this Perspective, however, we shall focus on a much less beaten track, namely
the use of quantum computers for the simulation of classical fluids
\footnote{Despite their early appearance, we shall not discuss the so-called type-II 
quantum computers \cite{YEPEZ1}, since they do not appear
to be universal}.
To this purpose, let us refer to a physics-computing 
plane defined by the following four-quadrants:
\begin{itemize}
\item[] CC: {\it Classical computing for Classical physics};
\item[] CQ: {\it Classical computing for Quantum physics};
\item[] QC: {\it Quantum computing for Classical physics};
\item[] QQ: {\it Quantum computing for Quantum physics}.
\end{itemize}
as represented graphically in Figure \ref{fig_CCQQ}.

Feynman was probably referring to the CQ sector shown in Figure \ref{fig_CCQQ}, where one 
is often faced with exponential complexity barriers (in his Nobel speech, Walter Kohn defines the 
many-body Schr\"odinger wavefunction as "unphysical" precisely because of this reason \cite{KOHN}). 
The basic idea is that these barriers are unphysical because we are using the wrong 
formalism to ask the question that lie behind them. Unlike Kohn, Feynman takes no issue at 
the N-body Schr\"odinger equation, but simply observes that it is unphysical only if 
we insist in solving it on the CQ quadrant instead of the QQ one. 
In line with this observation, the simulation of quantum 
many-body systems has received most of the quantum computing
attention in the recent past \cite{PVC,GG}. 

In this Perspective, we shall focus on the much less explored off-diagonal QC quadrant, the 
natural question being whether the power of quantum computing can be brought to the benefit 
of classical physics as well, with specific focus on classical fluid dynamics.

Yet, there is another, subtler, motivation along the Kohn-Feynman 
discussion above: while it is undeniable that physics is quantum, it is no 
less true that physics has a very strong innate tendency to become classical 
at sufficiently large scales (high temperature). 
Such tendency, the major source of troubles for quantum computing,
remains only partially understood and therefore
the "foundational" question is whether at some point, classical 
computing takes over and cannot be beaten by any quantum algorithm.
This is just the opposite of the standard notion of 
"Quantum Advantage", hence we may dub it "Classical Advantage". 

Speaking of Classical Advantage, the starting observation is that many classical 
problems feature two major elephants in the  quantum computing room: non-linearity 
and non-hermicity (dissipation).
How does quantum computing deal with the two "elephants" above?
This is the main question addressed by this Perspective, with specific  
focus on classical fluids \cite{PRAMA}. 

\section{Challenges facing quantum computational fluid dynamics (QCFD)}
As mentioned in the introduction, realising the potential of quantum computing means 
leveraging distinctive features of quantum mechanics that, by definition, are not available 
on classical computers.  However, it also follows that it is precisely these specific features that 
expose major challenges in realising simulations with a quantum advantage.

The main quantum mechanical concepts spawning the potential benefit of quantum 
algorithms are {\it quantum superposition} and {\it quantum parallelism}. 
The quantum state in an $Q$-qubit coherent register can be described by the 
Schr\"odinger wave function, defined by $2^Q$ complex amplitudes for $2^Q$ states in superposition. 
The square of each of these amplitudes defines the probability of finding the system 
in the corresponding state after {\it quantum measurement}.
By encoding classical data in terms of these amplitudes an exponential saving in 
storage can be achieved when the number of qubits is compared to required number of classical bits. 

Let us illustrate the idea for the specific case of turbulent flow simulation. 
Turbulence features a $Re^3$ complexity, where the Reynolds number $Re$ represents 
the relative strength of convection (nonlinearity) versus dissipation. 
Most real life problems feature Reynolds numbers in the many-millions; for instance 
an airplane features $Re \sim 10^8$, implying $O(10^{24})$ floating-point operations per
simulation.This is basically the best one can afford on a nearly-ideal Exascale classical
computer. The simulation of regional atmospheric circulation flows takes us at least another
two decades above in the Reynolds number, hence totally out of reach for any
foreseeable classical computer \cite{WEATHER,PALMER}.
On the other hand, the minimum number of qubits $Q$ required 
to represent $Re^3$ complexity can be roughly estimated as
$2^Q = Re^3$, namely:
\begin{equation}
Q \sim 10 log_{10} Re
\end{equation} 
This simple estimate shows that $80$ qubits match the requirement of full-scale
airplane design, while $O(100)$ qubits would enable regional atmospheric models.
However, several key challenges stand in the way of this task.
First, quantum measurement needed to extract classical information collapses the quantum wave vector. 
Hence, to get classical values for each of the amplitudes, multiple realisations of the quantum 
state vector are needed with an associated set of measurements.
Second, in the quantum circuit model, the 'classical' information is not available to the 
quantum gate operations performed in the circuit. 
Specifically, gate operations can be conditional on the state of one or 
more control qubits, while specifying gate operations conditional on 
one or more of the complex amplitudes defining the wave function is impossible. 
Therefore, when classical data is encoded in terms of amplitudes, for example, a 
rotation angle in a quantum gate operation, this information is required at the 
time the circuit is compiled. 
During the execution of the circuit, this angle cannot be changed as a function 
of 'classical' data encoded in quantum amplitudes.
Third, In an algorithm involving multiple iterations or time-steps, the overhead 
associated with quantum measurements used to extract classical data and re-initialization 
of the quantum state for a next iteration, scale quadratically with the grid size
\cite{STAN} and consequently they severely tax the quantum CFD efficiency.
Associated with these challenges, it should be observed that the well-known HHL\cite{HHL} 
algorithm for linear system solution, assumes that the input and output data are
encoded in terms of quantum amplitudes without including
the cost of setting up the quantum state and extracting the classical solution.

With the exception of quantum measurement operations, quantum mechanical operators 
are unitary, linear and reversible. 
These operators are implemented using a series of unitary quantum gates (quantum circuit model).
If we further assume that the velocity field is represented in terms 
of {\it amplitude encoding}\cite{Brassard2002}, i.e. the components of velocity vector 
at all grid points are represented in terms of the amplitudes defining the 
wavefunction, then the {\it no-cloning theorem} prevents the use of 
(temporary) copies of any of these amplitudes. 
So, evaluation of $u^2$ or $u \frac{\partial u}{\partial x}$ cannot be performed 
by storing a temporary copy $temp=u$, to perform the computation 
of the value of $u^2$ as $u \times temp$. 
Also, for data encoded in terms of the complex amplitudes of the Schr\"odinger wavefunction, there 
is a need for this state vector to have a unit norm, since these amplitudes represent 
probabilities of states. 
This means that for an operator attempting to compute the squares of these 
complex amplitudes, the resulting state vector loses unitarity. 
So, even without the no-cloning theorem complicating such a step, this points to a 
further problem with computing non-linear terms. 
A similar argument runs for orthogonality of the eigenstates, since a nonlinear
propagator rotates the state by an angle which depends on the state itself.

Dissipation is also a concern, since it breaks hermicity of the quantum propagator
However, several ways out can be conceived, one of the most popular ones being
to augment the system with its hermitian conjugate, so that the doubled system
is hermitian by construction. 

To summarize, dealing with non-linear terms can be regarded as one of 
the main challenges for Quantum Computational Fluid Dynamics (QCFD).

\section{Hybrid quantum/classical approaches}

The challenges sketched above relating to {\it non-linearity}, {\it non-hermicity}, in combination with 
quantum circuit depth limitations imposed by NISQ-era hardware, have resulted 
in the fact that most of the existing work in QCFD is based on 
a {\it hybrid quantum/classical approach}, with the quantum processor 
performing computations for which efficient quantum algorithms exist, while 
the output is then passed on to classical hardware to perform further 
computational tasks not (yet) amenable to quantum algorithms. 

\begin{figure}
\centering
\includegraphics[scale=0.3]{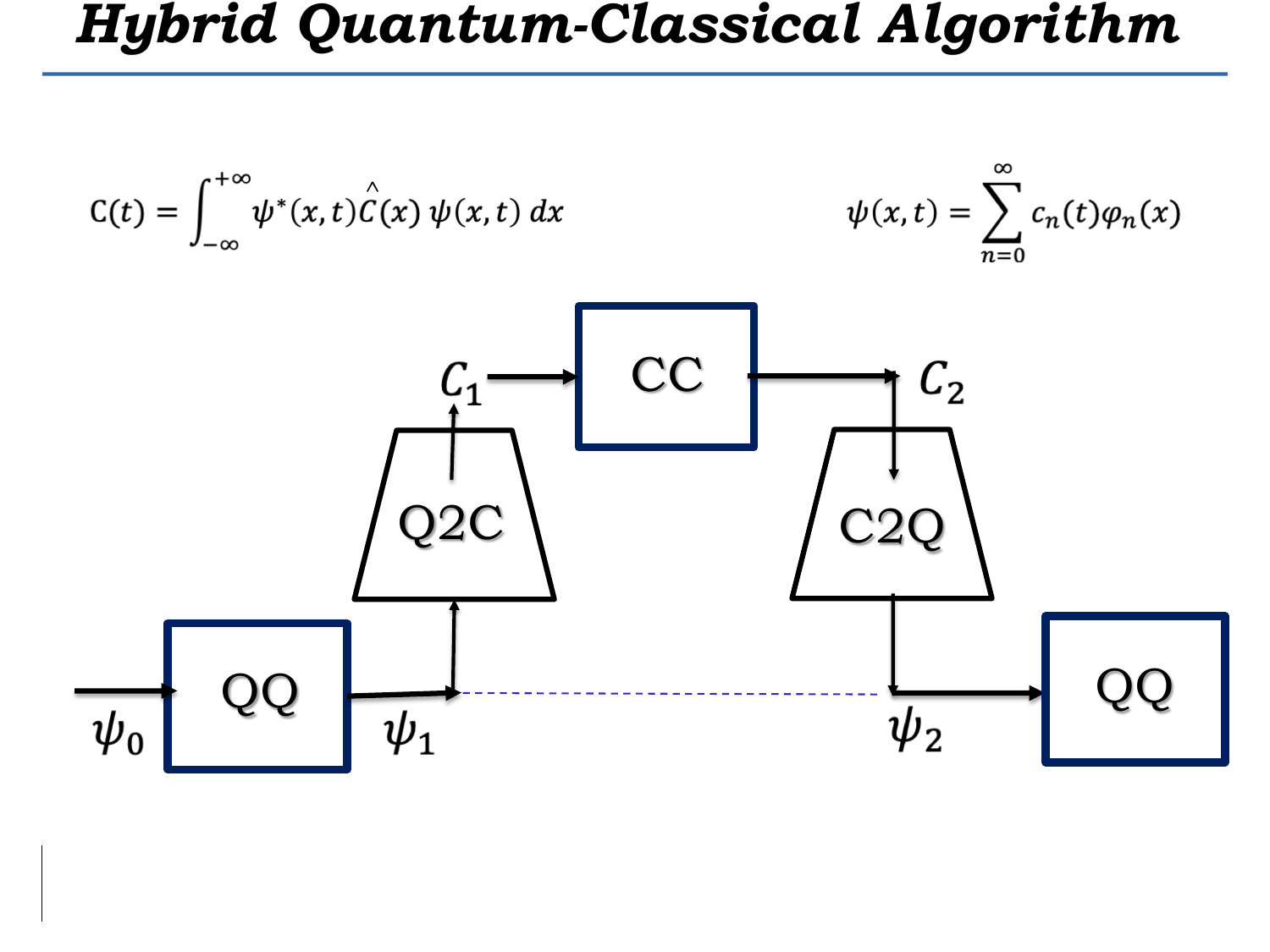}
\caption{Sketch of a hybrid quantum-classical algorithm. 
The illustration shows the steps involved in a single time step or single 
iteration, including preparation of the subsequent step/iteration.}
\label{fig_HybridQC}
\end{figure}

Figure \ref{fig_HybridQC} provides an illustration of the main 
concepts in this approach. As shown, quantum state $\psi_0$ is advanced to 
the next quantum state $\psi_1$ via a QQ algorithm. 
The quantum state $\psi_1$ is then used to generate 
classical observables $C_1$ which are advanced to $C_2$ by a CC algorithm.
The classical observables $C_2$ are then used to reconstruct the quantum 
state $\psi_2$, ready for the next QQ step. The Q2C conversion shown in Figure \ref{fig_HybridQC} involves quantum measurements and requires averaging over a statistical samples of quantum states, since
none of them can be reused. The C2Q reconstruction requires the preparation
of all the quantum eigenstates, hence a full reset of the quantum circuits from scratch. 
Both operations impose a substantial computational burden
on the hybrid algorithm. 
Specifically, the cost of initialization of an arbitrary quantum state with $Q$ qubits 
scales exponentially with $Q$. 
Initialization techniques with smaller overhead have been developed
for a limited set of specific quantum states. 

Examples of previous works using the {\it hybrid quantum/classical approach} include 
the works of Steijl\cite{RENE2019}, Gaitan\cite{GAITAN} and Budinski\cite{Budinski2022}. 
The algorithm presented by Gaitan uses Kacewiz's quantum amplitude estimation ODE 
algorithm \cite{KACEW} as applied to the set of nonlinear ODE's resulting from standard 
discretization of the Navier-Stokes equations. 
As shown, for certain 'non-smooth' problems (illustrated using the quasi-1D flow in 
converging-diverging duct with normal shock wave), the complexity analysis shows potential 
for exponential speed-up, so that the challenges associated with hybrid quantum/classical computing 
can potentially be overcome. 
For the linear advection-diffusion equation, Budinski\cite{Budinski2021} presented a quantum 
algorithm based on the Lattice Boltzmann method\cite{OUP1}. 
The algorithm can perform multiple successive time steps with no need 
for quantum measurements and re-initialization of qubit register between the time 
steps if suitable re-scaling of solution vector is applied to deal with non-unitarity. 
In the quantum circuit model, the fact that velocity field is unchanged between successive 
timestep means that an operator implementing $u \frac{\partial u }{\partial x}$ 
can be re-used in multiple time steps. 
Budinski\cite{Budinski2022} then extended the work to Navier-Stokes equations in 
streamfunction-vorticity formulation. Then, velocity-field updates during each time step means 
that quantum-circuit implementation of convection terms cannot be re-used during multiple 
timestep and that the 'classical' value of $u$ (as well as other flow field data) is needed 
to define quantum circuit implementation of for the next time step. 
This shows that it is the non-linearity that forces the use of a 
hybrid quantum/classical approach, similar to previous work where 
Navier-Stokes equations were discretized, e.g. the hybrid quantum/classical 
algorithms for fluid simulations based on quantum-Poisson solvers\cite{RENE2019}.

In summary, key challenges for the {\it hybrid quantum/classical approach} are:
i) Cost and complexity of (repeated) measurements;
ii) Statistical noise due to sampling of the quantum solution;
iii) Cost and complexity of (repeated) re-initialization.

\section{Quantum fluid dynamics strategies}
For teh sake of concreteness, let us
consider thw Navier-Stokes equations for time-dependent incompressible flows:
write as follows:

\label{sect_nonlinear}
\begin{eqnarray}
&&\frac{\partial {\bf u}}{\partial t}
+ {\bf u}\cdot\frac{\partial {\bf u}}{\partial {\bf x}}=
-\frac{\partial P}{\partial {\bf x}} + \nu \Delta {\bf u}
\label{eq_NS_mom}\\
&&\nabla \cdot {\bf u} = 0
\label{eq_NS_mass}
\end{eqnarray}
where ${\bf u}$ is the velocity vector, $P$ pressure, defined for location 
${\bf x}$ as function of time $t$. The kinematic viscosity (assumed independent of temperature) is 
defined by $\nu$ and density has been conventionally set to a unit constant value. 
Equation (\ref{eq_NS_mass}) enforces mass conservation, while Equation (\ref{eq_NS_mom}) is 
based on momentum conservation in each coordinate direction.
Equations (\ref{eq_NS_mom}) and (\ref{eq_NS_mass}) highlight that it is the convection term that represents 
the non-linearity, i.e. the second term on the left-hand side of Equation (\ref{eq_NS_mom}). 
Writing the Navier-Stokes equations in non-dimensional 
form, such that ${\bf x}$ and ${\bf u}$ are scaled by reference length $L_{ref}$ and $U_{ref}$, respectively, it 
follows that in Equation (\ref{eq_NS_mom}) the term $\nu$ is replaced by $1/Re$, where Reynolds 
number $Re=U_{ref} L_{ref}/\nu$. 
For Stokes flow, i.e. with Reynolds approaching $0$, nonlinear 
terms are vanishingly small, but still not to be neglected since they 
are responsible for non trivial long-range correlations especially 
important in biological flows \cite{RMP}. 
For high Reynolds number (turbulent) flows, obviously the nonlinear terms play the leading role.

In the following we present a cursory view of various existing strategies to simulate fluid dynamics
on quantum computers. 
\subsection{Nonlinear quantum ODE solvers}

A straight approach to quantum simulation of fluids consists in tackling the
non-linearity head-on, without trying to establish any parallel to quantum mechanics.
In this case, one would discretize the Navier-Stokes equations, turn them into set of 
nonlinear ODE's, to be solved by appropriate quantum nonlinear ODE solvers. 
Symbolically:
\begin{equation}
\frac{d {\bf u}}{dt} = f({\bf u})
\end{equation}
where ${\bf u}$ is a shortand for the full set of unknowns hosted by the computational grid.
A standard time marching scheme yields
\begin{equation}
{\bf u}(t+\Delta t) = {\bf u}(t) + \int_{t}^{t+\Delta t} f({\bf u}(z))dz
\end{equation}
and the quantum algorithm takes charge of performing the discrete summation
implementing the time-integration at the right hand side.

This approach has been pioneered by Gaitan \cite{GAITAN} for the case of a Laval
nozzle with encouraging results on grids with $O(30-60)$ grid points over $1400$ timesteps.
The dangling issue, though, is identification of an
appropriate quantum oracle for evaluating the right hand side $f(u)$, a task that
as already mentioned, Gaitan leaves to classical computers.   

\subsection{Nonlinear variational quantum eigenvalue solvers}

Always in the spirit of taking non-linearity head-on, but this time 
with a close eye to quantum physics, it has recently be proposed that 
variational quantum eigenvalue (VQE) solvers, a major tool of the QQ sector, might be extended
to the fluid equations as well \cite{CFDVQE}. The basic idea behind VQE is to use quantum 
computing to construct variational eigenfunctions and minimize the 
energy functional through a classical procedure.
In \cite{CFDVQE} the authors propose to use a similar technique for the Navier-Stokes
equations, i.e 1) construct variational trial functions and 2) minimize the associated
(dissipative) functional via a classical procedure.
Formally, step 1 consists in expressing the flow field in variational form
\begin{equation}
\label{VARU}
{\bf u}(x,t;\lambda) = \sum_n {\bf u}_n(t) \phi_n(x;\lambda)
\end{equation}
where $\phi_n$ is a suitable set of basis functions parametrically 
dependent on the a set of variational parameters $\lambda$.
Such variational parameters are then fixed by minimizing the energy 
dissipation functional $D(\lambda) = \nu \int (\nabla u(x;\lambda))^2 dx$
where the integral runs over the entire volume occupied by the fluid.
The appeal of this idea is twofold: first, it may borrow a lot of QQ know-how, second 
it bypasses the issue of quantum time marching.     
We are not aware of any practical implementation of the idea, but best guess is that
they will become available soon.

\subsection{Linearization: Carleman embedding}  
It is long known that any nonlinear problem can be mapped into a 
linear one in a space of higher dimensions, a technique
also known as Carleman embedding or Carleman linearization \cite{CARLE}.
Spatial discretization of the non-dimensional form of Equation (\ref{eq_NS_mom})
 using a spatial grid consisting of $l=1,G$ lattice sites, results in the 
following set of equations,
\begin{eqnarray}
\frac{\partial {\bf u}_l}{\partial t} = -{\bf u}_l\cdot\left (\frac{{ \partial \bf u}}
{\partial {\bf x}}\right)_l-\frac{1}{\rho}\left(\frac{\partial P}{\partial {\bf x}}\right)_l
+ \frac{1}{Re} \left(\Delta {\bf u}\right)_l
\label{eq_NS_ODE1}
\end{eqnarray}
where $()_l$ indicates that derivatives are evaluated at lattice site 
$l$ and ${\bf u}_l$ represents the velocity vector at lattice site $l$. 
Introducing indices $m$ and $n$ to denote interacting neighbors, and using Einstein summation, 
Equation (\ref{eq_NS_ODE1}) can be rewritten in terms of a linear operator 
$\mathcal{L}$ and quadratic operator $\mathcal{Q}$, as follows,
\begin{eqnarray}
\frac{d u_{l}}{dt} = \mathcal{L}_{lm} u_{lm} 
+ Re \; \mathcal{Q}_{lmn} u_{m} u_{n}
\label{eq_NS_ODE2}
\end{eqnarray}
The pressure term was removed for simplicity, although this all but a minor item, since
pressure-flow-stress coupling is going to heavily affect the structure of the Carleman matrix. 
At this stage, it should be noted that in the discretization of the velocity derivatives 
at lattice point $l$, the values of the velocity at one or more neighboring lattice points is used. 
This means that when the Carleman linearization is used to introduce a 
new variable $V_{lm} = u_l u_m$ to formally 
generate a linear system, marching the system of equations forward in time produces
an ever growing hierarchy in which the Carleman variables at level $k$ involve the Carleman 
variables at the next level $k+1$. 
Note that at each level we are faced with a tensor of rank $k$, 
with $O(G \kappa^{k-1})$ independent 
components, $\kappa$ being the sparsity of the $\mathcal{Q}$ matrix. 
Furthermore the tensors occupy a neighborhood of the original field
whose diameter grows linearly with the Carleman level.
This shows that there is a major prize to pay to uplift the 
nonlinearity of the fluid equations, both in terms of 
increasing dimensionality and loss of locality.

Nevertheless, in \cite{CHILDS} the authors  
present an algorithm based on Carleman linearization along with the 
use of a quantum linear system solution approach for the solution of the
one-dimensional Burgers equation. 
The presented algorithm shows a polylog scaling with the number 
of grid points, i.e. an exponential improvement over
classical approach. However, for a given time span $T$, the algorithm 
shows a complexity $T^2 Poly(logT)$, i.e. a significantly 
increased time- complexity compared to classical case. 
The authors simulate shock formation in a one-dimensional Burgers flow, with 
$16$ grid points over $4000$ time-steps, with a fourth order Carleman truncation.
They reach Reynolds numbers up to $40$, an order of magnitude
larger than predicted by the theoretical no-go analysis, a welcome discrepancy
which begs for further analysis.

The key question then is: how fast does the Carleman 
procedure converge as a function of the Reynolds number?

As we shall see, preliminary results offer room for optimism.

\subsection{Carleman Lattice Boltzmann}

The answer may depend on the chosen representation of the fluid 
equations, the Lattice Boltzmann (LB) method being a prominent candidate
in this respect.
In equations:
\begin{equation}
\partial_t f_i + {\bf v}_i \cdot \nabla f_i = - \frac{f_i-f_i^{eq}}{\tau}
\end{equation}
where $f({\bf x},{\bf v},t)=\sum_i f_i(x,t) \delta({\bf v} -{\bf v}_i)$, is the probability to find a
representative particle with discrete velocity ${\bf v}_i$ at position ${\bf x}$
and time $t$, $f_i^{eq}$ is the corresponding discrete local equilibrium
and $\tau$ is a local relaxation time, fixing the viscosity of the lattice fluid. 
Importantly, the local equilibrium is a quadratic function of the Mach number $Ma=u/c_s$,
$c_s$ being the speed of sound.

A key point of using the discrete-velocity Boltzmann formalism instead of Navier-Stokes 
is that, owing to the double dimensional phase-space, in the latter non-locality (streaming) 
is linear while non-linearity (collision) is local, while in Navier-Stokes the two merge into a single 
${\bf u} \nabla {\bf u}$ convective term. 
On classical computers this disentanglement proves extremely
beneficial and the idea is that similar benefits apply to the quantum case as well.
Indeed, most importantly for the quantum case, in the Boltzmann formulation the
nonlinearity is not measured by the Reynolds number, but by the Mach number instead,
which is typically well below 1, thereby dramatically lowering the nonlinearity barrier. 

Based on the Lattice Boltzmann method, Itani et al.\cite{WAEL} employ Carleman 
linearization to develop an approach termed \textit{Carleman for second-quantized Lattice Boltzmann}. 
This terminology stems from the fact that the Boltzmann operator, defined by:
$
\mathcal{B} f_i = - {\bf v}_i \cdot \nabla f_i - \frac{f_i-f_i^{eq}}{\tau},
$
can be expressed in terms of the second-quantization annihilation and generation 
operators via the relation $\nabla = (\hat a-\hat a^+)$. 
As a result, the formal solution $f_t = e^{\mathcal{B}t} f_0$ can be 
computed in close analogy with quantum mechanics.

The quantum computing algorithm for streaming is based on the approach used 
by Steijl and co-workers\cite{RENE2020}, while collisions follow the 
bosonic encoding first proposed by Mezzacapo et al, whereby dissipative
effects are repriesented as a weighted sum of two unitary operators \cite{MEZZA}. 
The scheme is as "Feynmansque" as it can possibly get, as it builds on 
a one-to-one analogy between LB and the Dirac equation, first proposed in \cite{QLB}.
However, it is  subject to a number of additional questions: primarily 
truncation effects due to the finite number of bosonic excitations and 
the long-time behaviour of non-unitarity errors \cite{WAEL23}.

Recently, Cheung and coworkers performed a Taylor-Green
vortex simulation based on a Carleman-LB scheme, 
showing excellent agreement at moderate
Mach number, with just three Carleman iterations \cite{MARGIE}. Although much more
work is needed to consolidate these preliminary results, there appears to be room 
for some optimism on the use of Lattice-Boltzmann-Carleman quantum computing schemes.

\subsection{Functional Liouville Approach}
The Carleman embedding provides a linearization of the actual equations of fluid motion. 
An alternative procedure is to take the statistical dynamics approach, whereby 
one seeks the probability distribution  function (PDF) of the fluid velocity field
\cite{DSFD22}. 
This is formally straightforward since the PDF obeys a functional continuity 
equation, best know as Liouville equation:
\begin{equation}
\partial_t P[{\bf u}] + 
\frac{\delta}{\delta {\bf u}} 
(f({\bf u}) P[ {\bf u}])=0
\label{LIU}
\end{equation}
where $P=P[{\bf u}]$ is the functional PDF and $\dot {\bf u} = f({\bf u})$ is the nonlinear 
equation of motion (hence $f({\bf u})$ is an operator in function space). 
Note that regardless of the nonlinearity of the dynamics, reflected 
by $f({\bf u})$, the functional equation is linear, by construction.
Once the fluid equations are discretized on a grid with $G$ lattice nodes, the Liouville 
distribution $P_G$ becomes a $G$-variate PDF $P_G({\bf u}_1 \dots {\bf u}_G)$
which lives in a $O(G)$-dimensional space, where 
for large-scale turbulent flow applications, $G$ can reach values of multi-billions. 
However, it should be noted that for most practical purposes, 
the $G$-body Liouville PDF is an hotel with vastly more rooms than customers,
meaning that (much) lower order marginals often suffice to deliver 
the essential physical information.
Key to the success of the program is the ability to find an appropriate 
closure, namely an effective kinetic equation for the low order marginals.
This is a classical topic in non-equilibrium statistical physics, which 
may draw significant benefits from modern developments in 
tensor networks theory \cite{TN,TN2}. 

In \cite{Jin2023}, the authors develop a general and elegant framework based on the
Koopman-von Neumann and the level-set  approach to classical nonlinear 
field theories \cite{Jin2023}. 
The formalism is applied to hyperbolic PDE's and Hamilton-Jacobi equations,
but its extension to the Navier-Stokes equations is addressed only 
marginallyi, making it difficult to draw firm conclusions.

\subsection{Quantum-Fluid Duality: Inverse Madelung Transform}

Formal analogies between quantum mechanics and fluid dynamics have been noted since
the early days of quantum physics, most notably with the work of Madelung, who noted
that upon writing the wavefunction in eikonal form $\Psi = \rho^{1/2} e^{i \theta}$, the
imaginary part of the Schroedinger equation turns into the continuity equation, while
the real one provides the follow madelung fluid equation
\begin{equation}
\label{MADEL}
\partial_t {\bf u} + {\bf u} \cdot \nabla {\bf u} = - \nabla (V_C+V_Q) 
\end{equation}
where ${\bf u}=(\hbar/m) \nabla \theta$ is the fluid velocity, $V_C$ the classical
potential and 
$V_Q= (\hbar^2/2m) \rho^{-1/2} \Delta \rho^{1/2}$ is the so-called quantum potential.
The Madelung formulation shows that the Schroedinger equation behaves like an inviscid,
irrotational fluid subject to the classical potential $V_C$ plus a genuinely
quantum potential $V_Q$.
This analogy has interesting interpretations for the hidden-variables 
theory of quantum mechanics, which are beyond scope of this Perspective. 
Here we simply wish to highlight its operational value,
i.e. it permits to solve quantum problems using numerical methods for fluids.
In the context of quantum computing, we are interested in taking the opposite path, namely
using quantum computers to solve fluid problems in quantum mechanical vests.
This is a fascinating possibility, with the major caveat that the Madelung fluid is a far distant
relative of classical Navier-Stokes fluids, the main points of departure being i) dissipation, 
ii) non-gradient flow, iii) absence of quantum potential.
Very recent work has shown how to mend points ii) and iii) using quaternions, but still leaves 
item i) open \cite{MADINV}.
It is of decided interest to explore whether a generalized Schroedinger equation
matching exactly Navier-Stokes fluids can be found and efficiently simulated on quantum computers. 
\begin{figure}
\centering
\includegraphics[scale=0.3]{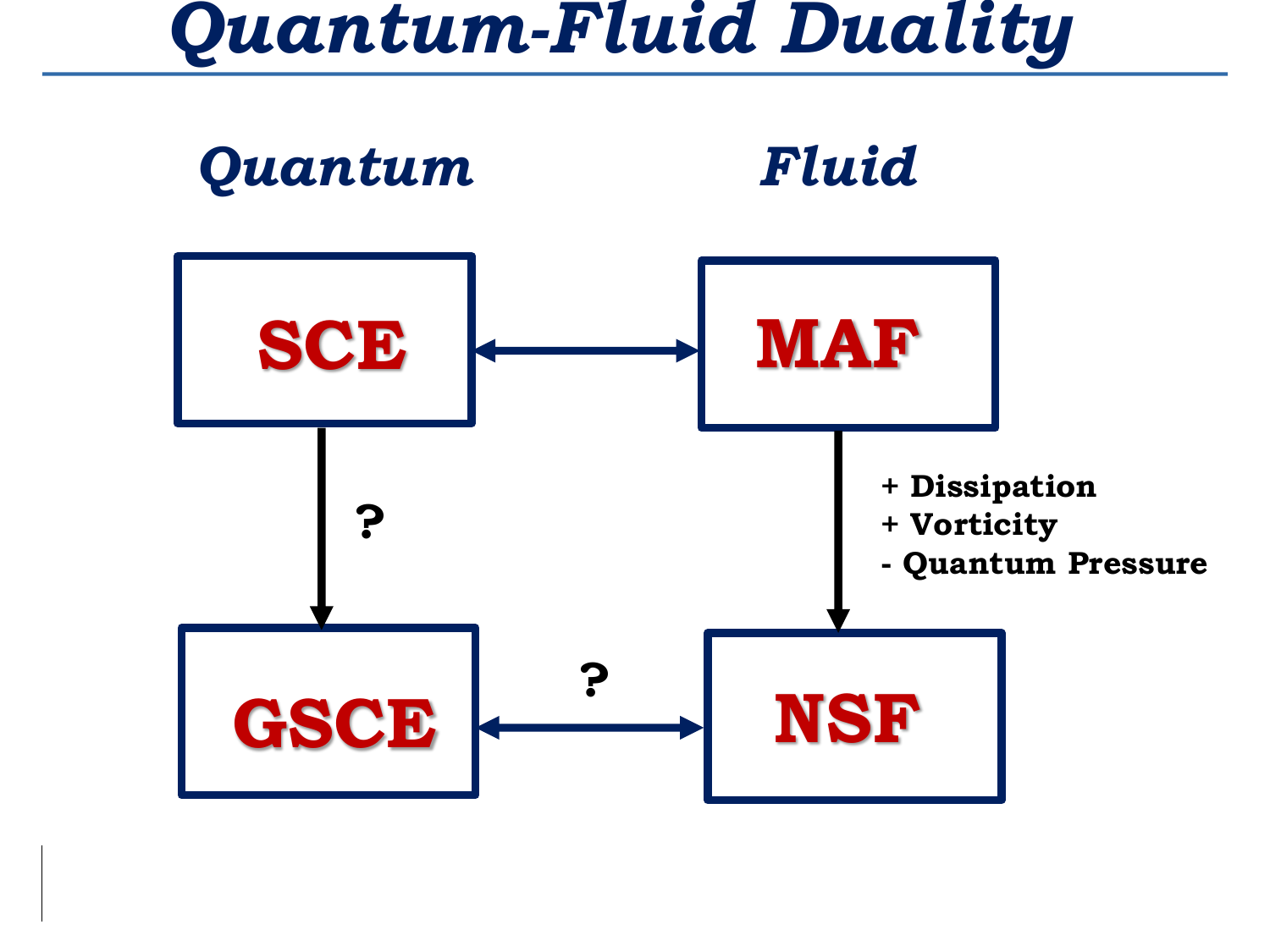}
\caption{Schematics of the quantum-fluid duality.
The Schroedinger equation (SCE) maps one to one to the Madelung fluid 
(MAF), which is compressible, inviscid, irrotational and subject 
to quantum potential (pressure).
The idea of the inverse Madelung transform is to generate a generalized 
Schroedinger equation (GSCE) equivalent to the compressible Navier-Stokes fluid. 
By solving the GSCE on a quantum computer one would then solve the Navier-Stokes physics. 
}
\label{Madelung}
\end{figure}
\section{Low-Reynolds flows}

Last but definitely not least, the physics of fluids is littered  with interesting  
problems at low-Reynolds, especially in microfluidics, soft matter and biology 
\cite{STONE,MICRO,OUP2}.
For instance, it would be of great interest to devise a 
{\it quantum multi-scale} application, coupling quantum algorithms for 
biomolecules swimming in a water solvent described by a quantum algorithm for
low-Reynolds fluid flow. Given that low-Reynolds flows are nonlocal, perhaps
the inherent nonlocality of quantum mechanics could prove helpful in
representing the classical nonlocality of low-Reynolds flows.

\section{Conclusions}

In summary, we have presented a survey of the main current approaches to 
the quantum simulation of classical fluids. 
Various obstacles stand in the way of the efficient simulation of fluid flows
on quantum computers, especially at high-Reynolds numbers, mostly on 
account of the strong nonlinear effects.
A few ways potential ways out have been illustrated, but their practical 
implementation commands major advances in quantum technology \cite{DAS}.  

On philosophical grounds, it can't be ruled out that high Reynolds
flows would show "classical advantage", namely no quantum algorithm can 
ever beat a classical one beyond a given threshold of nonlinearity.
Indeed, as observed earlier on, while it is true that Nature isn't classical, it 
is equally true that Nature has a very strong innate tendency to {\it become}
classical at macroscopic scales/high temperatures. 
Assessing to what extent quantum computers can withstand such 
tendency, or possibly even take advantage of it, is not only of 
practical but also of major foundational interest. 
Quantum computers offer indeed a unique opportunity to ask questions that
the founding fathers of quantum mechanics could only formulate as "Gedanken Experiments".

\acknowledgments

The authors have benefited from valuable discussions 
with many colleagues, particularly
S.S. Bharadwaj, D. Buaria,
P. Coveney, N. Defenu, 
G. Galli, M. Grossi, B. Huang, A. Mezzacapo, 
S. Ruffo, A. Solfanelli and T. Weaving.
S.S. wishes to acknowledge enlightening discussion with M. Berry.
He also acknowledges financial support form the Italian National
Centre for HPC, Big Data and Quantum Computing (CN00000013).

\end{document}